\documentclass{article}

\usepackage{arxiv}

\usepackage[utf8]{inputenc} 
\usepackage[T1]{fontenc}    
\usepackage{hyperref}       
\usepackage{url}            
\usepackage{booktabs}       
\usepackage{amsfonts}       
\usepackage{nicefrac}       
\usepackage{microtype}      
\usepackage{lipsum}
\usepackage{graphicx}

\title{Detectability of Covert Fissile \\ Material Production in Nuclear Fusion Reactors \\ via Antineutrino Emissions}

\author{
 Alexander Glaser \\
  Department of Mechanical and Aerospace Engineering \\
  Princeton University \\
  Olden St, Princeton, NJ 08542 \\
  \texttt{alx@princeton.edu} \\
  \And
 Robert J. Goldston \\
  Princeton Plasma Physics Laboratory \\
  Princeton University \\
  100 Stellarator Rd, Princeton, NJ 08540 \\
  \texttt{goldston@pppl.gov} \\
  \And
 Patrick Huber \\
  Center for Neutrino Physics \\ 
  Department of Physics \\
  Virginia Polytechnic Institute and State University \\
  Blacksburg, VA 24061 \\
  \texttt{pahuber@vt.edu} \\
}

\begin{document}
\maketitle
\begin{abstract}
Fusion power systems can in principle be used to make significant amounts of fissile material. To do so, an operator would have to introduce fertile material, such as uranium-238, in a suitable region of the reactor where it is exposed to an intense neutron flux. The possibility of using a fusion reactor for this purpose has raised the question of how these facilities can be monitored to ensure their peaceful use. This study examines whether covert production of fissile material in a declared fusion plant could be detected with an onsite antineutrino detector. We find that even a relatively small detector should be able to confirm production rates of a few kilograms of plutonium over 30 days, despite the cosmogenic background and the antineutrino emissions associated with neutron activation of reactor components.
\end{abstract}


\section{Introduction}

Nuclear fusion is experiencing a resurgence of interest, and growing research and development efforts may offer a path toward commercial deployment by mid-century. Fusion would provide carbon-free electricity and benefit from an inexhaustible supply of deuterium fuel and abundant lithium resources. Numerous concepts are currently under development \cite{gol24}; most of these concepts are based on the deuterium-tritium (DT) fusion process:

\[ {^2_1}\mbox{D} + {^3_1}\mbox{T}  \longrightarrow {^4_2}\mbox{He} + {^1_0}n + 17.6\,\mbox{MeV} \]

The tritium component of the fuel is produced in-situ during operation of the reactor, primarily after neutron absorption in lithium-6 via:

\[{^6_3}\mbox{Li} + {^1_0}n \longrightarrow {^3_1}\mbox{T} + {^4_2}\mbox{He} + 4.8\,\mbox{MeV} \]

Standard operation of a fusion reactor does not involve nuclear materials, which offers significant nonproliferation benefits compared to nuclear fission reactors. The presence of intense neutron fluxes provides an environment, however, that could be used for covert production of fissile materials. Such a production scenario could be realized by adding relatively small amounts of fertile material (uranium-238 or thorium-232) into a blanket, which surrounds the reacting DT fuel, extracts heat and produces tritium via neutron capture. We previously considered the scenario where small uranium-bearing particles are injected into the liquid lead-lithium coolant of a fusion reactor \cite{gla12}. These particles could be mechanically removed after exposure to fusion neutrons and the fissile material (plutonium or uranium-233) then be extracted from the matrix. When operated in such a mode, a gigawatt-scale fusion reactor could in principle produce tens of kilograms of plutonium or uranium-233 per week. The possibility of using a fusion reactor for this purpose has raised the question of how these facilities can be monitored to ensure their peaceful use with minimum operational impact. The fact that no nuclear material ought to be present at the site should simplify the inspection task significantly. In particular, any attempt to breed fissile material inevitably also results in concurrent fission events in the nuclear material during neutron exposure, especially in the presence of high-energy neutrons from DT fusion reactions.

This study examines whether covert production of fissile material could be detectable with an onsite antineutrino detector.  This could in turn inform efforts to develop effective monitoring approaches for future fusion reactors.


\section{Methods}

As part of this analysis, we model a simplified toroidal fusion reactor with a major radius of 6.2~meters and a minor radius of 2.0~meters using the Monte Carlo particle transport code MCNP6 \cite{mcnp}. The reference reactor operates at a typical power density and a total plasma fusion power of 1500\,MW using the DT reaction (17.6\,MeV per fusion event, including 14.1\,MeV carried by the neutron). The analysis below examines two alternative blanket designs illustrated in Figure 1: a molten salt (LiF-BeF$_2$ or FLiBe) design and a dual-coolant lead-lithium (DCLL) design \cite{bal25,rap21}. 
Irrespective of the specific blanket type, both reactor designs share several common structural elements, including a plasma-facing component, the first wall, breeding channels, and shielding regions. 
In our model, the plasma facing component consists of a thin layer of metallic tungsten. Reduced activation ferritic-martensitic steel (F82H) is used for structural materials. Neutron activation will be relevant in all major regions of the blanket, including in the design-specific breeding channels.

The neutronics of the reactor strongly depend on the lithium-6 enrichment used in the blanket. As part of this analysis, we explored several enrichment levels to identify reference designs with an adequate tritium breeding ratio on the order of 1.2--1.3.
Based on the results of these scoping studies, we selected a DCLL-blanket design that uses lithium enriched to 90\% in the isotope lithium-6 to maximize tritium production. Very high lithium enrichment is less critical for FLiBe blankets, and we use a value of 20\% for the analysis below; above this value, the tritium-breeding ratio no longer increases, which is consistent with previous findings \cite{seg20}.

\begin{figure}[hbt] \centering \sf
\includegraphics[width=130mm]{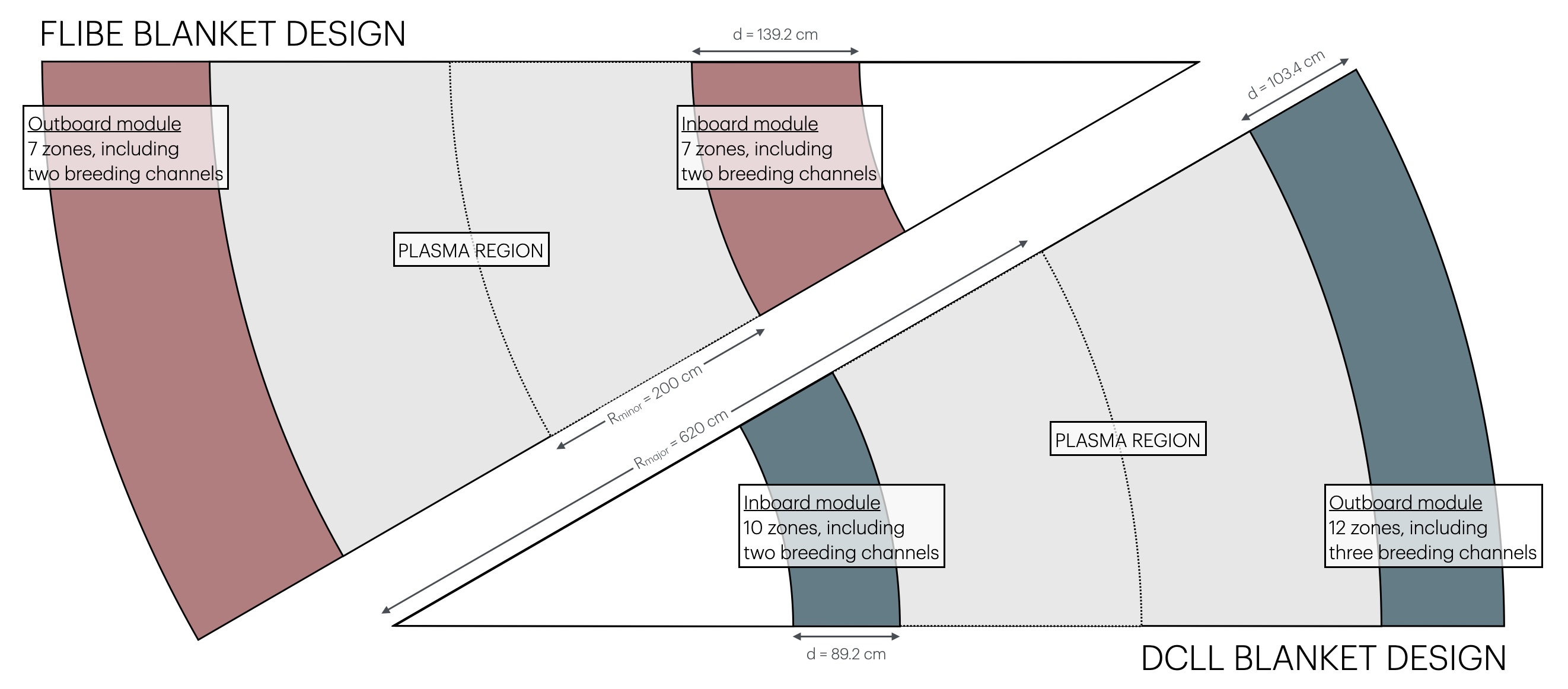}
\caption{Geometries of the molten-salt (FLiBe) design and the dual-coolant lead-lithium (DCLL) design simulated in MCNP6. Toroidal geometries are approximated using equivalent cylindrical models (30-degree segments shown here). Blankets consist of several radial zones, including breeding channels, where tritium is produced and where fertile material could be injected.}
\label{fig:design}
\end{figure}

The Monte Carlo transport code MCNP6 is used to determine all fundamental quantities needed for subsequent fuel depletion, including tritium breeding, and neutron activation analysis. Quantities estimated with MCNP6 include the radial neutron flux distribution in the blankets, neutron energy spectra, energy deposition, and reaction rates for all relevant reaction types and isotopes present in the different zones of the reactor.

These results, calculated at the beginning-of-life of the reactor, are then used in a separate step to estimate fissile material production rates and neutron activation and associated antineutrino emissions over time.

To enable this analysis, we developed a small suite of software tools that solve the respective depletion and activation equations for all regions of the reactor based on the data calculated with MCNP6. The software tracks the activation products from neutron captures and $(n,2n)$ reactions for all isotopes present in the reference design. It currently includes 27 stable parent isotopes and 30 radioactive daughter isotopes. The list of isotopes and production pathways can easily be modified or expanded as other reactor materials are considered.


\section{Results}

To conduct the analysis, we proceed in three main steps. We first define and evaluate a reference fissile material production scenario, estimating both plutonium production rates and the associated energy release as some of the covertly introduced nuclear material undergoes fission. This analysis provides the first part of the antineutrino background signal. We then determine neutron activation and antineutrino signatures driven by neutrons emanating from the fusion plasma, which provides the second part of the background. Finally, we use these signatures to assess the detectability of covert fissile material production against the combined background due to neutron activation and cosmogenic antineutrinos.

\subsection{Covert Fissile Material Production Scenario}

Before estimating antineutrino emissions during routine operation of the reactor, we define a reference scenario for covert fissile material production. The challenge will be to detect antineutrino emissions from fission events against the inevitable background of antineutrino emissions from neutron activation.

In the scenario considered here, blankets are ``seeded'' with millimeter-sized uranium-bearing particles \footnote{These uranium-bearing particles are inspired by tri-structural isotropic (TRISO) fuel, originally developed for high-temperature fission reactors. We use a simplified particle design with two rather than three layers of coating surrounding a spherical uranium kernel. This bi-isotropic (BISO) fuel is preferable for breeding of fissile material because it has a higher volume fraction of uranium compared to the more common TRISO fuel. \textit{High Temperature Gas Cooled Reactor Fuels and Materials}, Tech. Rep. IAEA-TECDOC-1645 (International Atomic Energy Agency, Vienna, Austria, 2010).}, which would have to be tailored for suspension in the carrier material \footnote{We assume that mechanical removal of particles would be (far) easier to implement, especially surreptitiously, than extraction of a dissolved compound, which would most likely require specialized equipment and processes.}. In our reference scenario, the particles contain natural uranium (0.72\,at\% U-235), rather than depleted uranium or thorium. In the MCNP6 simulations, the blanket material is ultimately homogenized while preserving the respective isotopic fractions in the material of the blanket zone.

\begin{figure}[hbt] \centering 
\includegraphics[width=110mm]{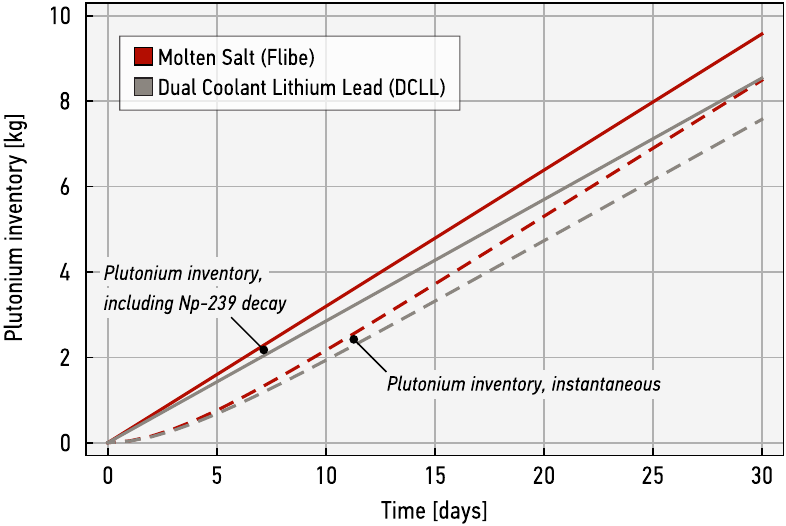}
\caption{Plutonium buildup in the blanket. Even for a relatively low concentration of 25 uranium-bearing particles per cubic centimeter, both blanket designs yield about one significant quantity (8\,kg) of plutonium in one month. Dashed lines exclude the contribution of neptunium-239, which decays into plutonium-239 with a half-life of 2.4 days. Calculations are based on spectrum-averaged cross sections and neutron flux values determined in MCNP6 simulations for the reference reactor operating at 1500 MW of fusion power.}
\label{fig:plutonium}
\end{figure}

We examined a range of particle loadings, but focus the misuse case on a relatively low density of 25 particles per cubic centimeter of blanket material, which corresponds to a volume fraction of about 1.3\%. This concentration results in a plutonium breeding rate that is far lower than theoretically possible, i.e., it provides a more challenging scenario for detection. As shown in Figure~\ref{fig:plutonium}, both blanket designs yield on the order of eight kilograms of plutonium over a one-month period: the totals add up to 9.6 kg for FLiBe and 8.5 kg of plutonium for DCLL when the contributions from delayed neptunium-239 decay are factored in.
The reduction of the tritium breeding ratio for this fissile material production scenario is very small ($\sim$\,1\%) and unlikely to affect sustained operation of the plant.

MCNP6 simulations show that the fission power generated in this scenario is about 60\,MW for the FLiBe blanket and about 85\,MW for the DCLL blanket, i.e., fission events do not contribute more than about 6\% of the power generated. It is reasonable to assume that this extraneous power could be safely handled by the cooling system; otherwise, the operator could lower the plasma power by a proportionate amount.

For the scenarios considered here, about 90\% of the fission power is due to fast-neutron fission of uranium-238. The contribution of uranium-235 is on the order of 10\%, and the contribution of plutonium-239 is negligible, even after several months of exposure. Using depleted uranium in the particles would therefore not significantly reduce fission power generation in the system.

Below, we examine whether a production rate of about eight kilograms of plutonium per month would be detectable from the associated antineutrino emissions and what the requirements for the detector system would be.


\subsection{Neutron Activation}

About $5.32\times10^{20}$ neutrons per second emerge from the plasma of a reactor using DT fuel and operating at a fusion power of 1500 MW. These neutrons impinge on the plasma facing component and first wall, before entering deeper into the blanket, where their kinetic energy is recovered. While neutron absorption in lithium is desirable to ensure sufficient tritium production rates, neutrons also interact with most other materials present. In particular, neutron capture and $(n,2n)$ reactions result in activation of the structural components, the blanket itself, and other parts of the reactor. Major R\&D efforts are underway to develop materials that minimize neutron activation while maintaining adequate irradiation performance under the extreme conditions expected for a commercial fusion reactor.

We determine the buildup of short-lived radionuclides in all regions of the reactor as a function of time with the custom software tools using neutron flux distributions and reaction rates calculated with MCNP6. Based on their inventories, the beta activities of the isotopes of interest can be determined via $A_i(t) = \lambda_i \, N_i(t)$. 

Tables~\ref{tab:flibe-activation} and \ref{tab:dcll-activation} summarize key results for the FLiBe and DCLL blankets, including total activities after 30 days and after 360 days for the most important isotopes present in the system.

\begin{table}[hbt] \centering
\begin{tabular}{cccccc}
\hline
\textrm{Isotope} &
\textrm{Decay} &
\textrm{Q [MeV]} &
\textrm{$A_{30}$  [Bq]} &
\textrm{$A_{360}$ [Bq]} & \\
\hline \hline
 W-181 & $\epsilon$ &  0.205 & 6.97E+18 & 3.83E+19 & \\
 W-185 & $\beta^-$  &  0.431 & 5.31E+18 & 2.11E+19 & \\
 W-179 & $\epsilon$ &  1.062 & 1.41E+19 & 1.40E+19 & \\
  F-18 & $\beta^+$  &  0.634 & 1.27E+19 & 1.27E+19 & \\
 W-187 & $\beta^-$  &  1.313 & 6.36E+18 & 6.32E+18 & \\
 Cr-51 & $\epsilon$ &  0.752 & 2.61E+18 & 4.93E+18 & \\
  F-20 & $\beta^-$  &  7.024 & 3.28E+18 & 3.28E+18 & $\leftarrow$ \\
  V-49 & $\epsilon$ &  0.602 & 2.52E+16 & 2.19E+17 & \\
  Li-8 & $\beta^-$  & 
         \hspace{-1ex}16.004 & 1.64E+17 & 1.64E+17 & $\leftarrow$ \\
 Cr-49 & $\beta^+$  &  1.608 & 7.81E+16 & 7.81E+16 & \\
 Ti-45 & $\beta^+$  &  1.040 & 1.25E+16 & 1.25E+16 & \\
 Fe-55 & $\epsilon$ &  0.231 & 1.06E+15 & 1.13E+16 & \\
 Ti-51 & $\beta^-$  &  2.470 & 6.64E+15 & 6.64E+15 & $\leftarrow$ \\
 Cr-55 & $\beta^-$  &  2.602 & 5.96E+15 & 5.95E+15 & $\leftarrow$ \\
Ta-182 & $\beta^-$  &  1.816 & 6.64E+14 & 3.54E+15 & $\leftarrow$ \\
 Mn-56 & $\beta^-$  &  3.695 & 2.11E+15 & 2.11E+15 & $\leftarrow$ \\
\hline
\end{tabular}
\medskip
\caption{\label{tab:flibe-activation}%
Neutrino emission rates from the reference reactor using the FLiBe blanket. Shown are values after 30 days ($A_{30}$) and after 360 days ($A_{360}$). Only anti-neutrinos from $\beta^-$ decay and above the threshold of 1.806\,MeV are relevant for detection using inverse beta decay; these entries are highlighted. All other reactions and events will be invisible to the detector.}
\end{table}

\begin{table}[hbt] \centering
\begin{tabular}{ccccc@{}c}
\hline
\textrm{Isotope} &
\textrm{Decay} &
\textrm{Q [MeV]} &
\textrm{$A_{30}$  [Bq]} &
\textrm{$A_{360}$ [Bq]} & \\
\hline
Pb-203 & $\epsilon$ & 0.975 & 1.09E+20 & 1.09E+20 & \\
 W-181 & $\epsilon$ & 0.205 & 4.67E+18 & 2.57E+19 & \\
 Fe-55 & $\epsilon$ & 0.231 & 2.30E+18 & 2.47E+19 & \\
 W-185 & $\beta^-$  & 0.431 & 5.75E+18 & 2.28E+19 & \\
 W-179 & $\epsilon$ & 1.062 & 1.40E+19 & 1.39E+19 & \\
 W-187 & $\beta^-$  & 1.313 & 7.81E+18 & 7.77E+18 & \\
Pb-209 & $\beta^-$  & 0.644 & 2.95E+18 & 2.95E+18 & \\
 Cr-51 & $\epsilon$ & 0.752 & 1.02E+18 & 1.94E+18 & \\
Ta-182 & $\beta^-$  & 1.816 & 1.48E+17 & 7.89E+17 & $\leftarrow$ \\
 Mn-56 & $\beta^-$  & 3.695 & 4.59E+17 & 4.59E+17 & $\leftarrow$ \\
 Fe-53 & $\beta^+$  & 2.721 & 3.74E+17 & 3.73E+17 & \\
 Si-31 & $\beta^-$  & 1.492 & 3.50E+17 & 3.50E+17 & \\
  V-52 & $\beta^-$  & 3.976 & 2.63E+17 & 2.63E+17 & $\leftarrow$ \\
 Fe-59 & $\beta^-$  & 1.565 & 8.20E+16 & 2.19E+17 & \\
 Mn-54 & $\epsilon$ & 1.377 & 7.57E+15 & 6.46E+16 & \\
 Cr-55 & $\beta^-$  & 2.602 & 4.12E+16 & 4.11E+16 & $\leftarrow$ \\
\hline
\end{tabular}
\medskip
\caption{\label{tab:dcll-activation}%
Neutrino emission rates from the reference reactor using the DCLL blanket. Shown are values after 30 days ($A_{30}$) and after 360 days ($A_{360}$). Only anti-neutrinos from $\beta^-$ decay and above the threshold of 1.806\,MeV are relevant for detection using inverse beta decay; these entries are highlighted. All other reactions and events will be invisible to the detector.}
\end{table}

In the case of the FLiBe blanket, after one year of operation, radioactive decays involving neutrino emissions following neutron capture or $(n,2n)$ events add up to about 20\% of the neutron emission rate from the plasma. Emissions from the DCLL blanket are even higher due to a particular strong contribution from lead-203.

As the tables indicate, absolute neutrino emissions are dominated by a relatively small number of isotopes. In the blanket designs considered here, unstable isotopes of tungsten are very prominent, even though the material is only used in a thin layer as part of the plasma facing component. Tungsten isotopes do not produce a relevant background, however, because they either decay via electron capture or have Q-values below the 1.806 MeV threshold. Similarly, in the case of the DCLL blanket, neutrino emissions following electron capture in lead-203 do not contribute to our background. The next section discusses the antineutrino signatures that are most relevant for the detectability of covert fissile material production in a DT fusion reactor.


\subsection{Antineutrino Signatures}

Antineutrinos were discovered in 1956~\cite{Cowan:1956rrn} using the P-Reactor at the Savannah River Site as a source and using the inverse beta-decay (IBD) reaction on protons $p + \bar\nu_e \rightarrow n + e^+$; this reaction remains the main detection method for antineutrinos today.  Given the technological maturity of IBD-based neutrino detectors we investigate whether they could be a suitable tool to detect the presence of fission and hence the production of fissile material.

\begin{figure}[bht] \centering 
\includegraphics[width=110mm]{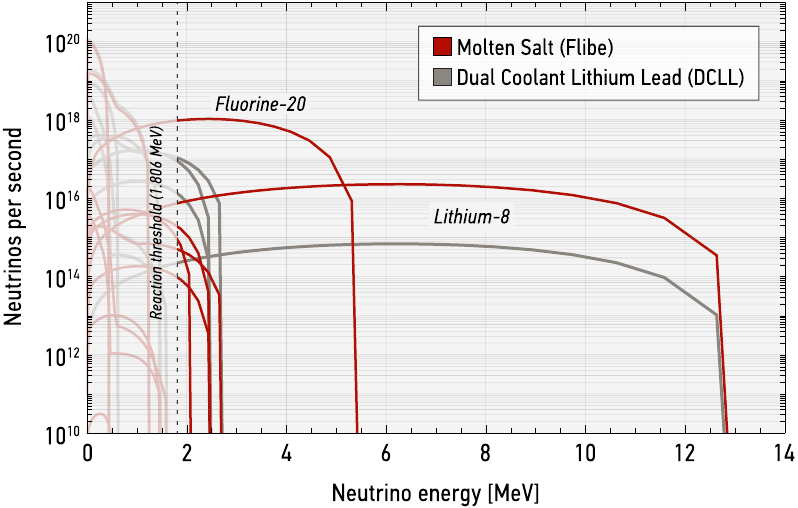}
\caption{Antineutrino emissions from a fusion reactor from activation products in the blanket. Dashed lines are for the FLiBe design, whereas solid lines are for the DCLL design. The gray shaded region indicates energies invisible to inverse beta decay. For both designs we assume 1500\,MW of fusion power.}
\label{fig:emission_spectrum}
\end{figure}

The IBD reaction has an energy threshold of 1.806\,MeV and can only occur for electron antineutrinos. Therefore, any beta decays with Q-values below that threshold and all beta-plus decays (producing electron neutrinos) are invisible to this reaction channel. The outgoing positron carries the energy of the incoming neutrino minus 1.806\,MeV and thus measuring its energy provides a direct measurement of the antineutrino energy; the cross section is well known~\cite{Vogel:1999zy}. The spatial and temporal coincidence of the resulting neutron and positron are the key to suppress non-neutrino backgrounds arising from cosmic rays and natural radioactivity. The neutron activation of material in the blanket will create an irreducible antineutrino background and we find that the leading contributors from neutron activation are lithium-8 and, in the FLiBe case, fluorine-20. Additional contributions from other isotopes are much smaller and practically irrelevant; see Fig.~\ref{fig:emission_spectrum} and Tabs.~\ref{tab:flibe-activation} and \ref{tab:dcll-activation}.  

The antineutrino yield from fission depends on the energy of the neutrons causing the fission since the resulting distribution of fission fragments changes with the neutron energy. The antineutrino yield from thermal neutron fission of both uranium-235 and plutonium-239 is well-studied and understood, both experimentally~\cite{DayaBay:2025ngb} and theoretically~\cite{Huber:2011wv,Kopeikin:2021ugh,Perisse:2023efm}. In a fusion reactor, however, most of the fission would be on uranium-238 and induced by 14\,MeV neutrons. There is limited experimental data on uranium-238 antineutrino yields from fission by fission-spectrum neutrons~\cite{Haag:2013raa}, and there is no data on fission by 14\,MeV neutrons. We performed a summation calculation of the antineutrino yield with the CONFLUX package~\cite{Zhang:2025ayc} using the cumulative fission yields for 14\,MeV neutrons~\cite{plompen_joint_2020}. We find that the resulting IBD cross section per fission~\footnote{The integral with respect to energy of the cross section and antineutrino yield.} is $7.5\times10^{-43}\,\mathrm{cm}^2$ compared to the measured value of $(8.51 \pm 0.95)\times10^{-43}\,\mathrm{cm}^2$ \cite{Haag:2013raa} obtained for fission neutron energies. The difference between our calculated value and the measured one is due the different neutron energies causing the fission and consistent with similar results on other fissile isotopes~\cite{Littlejohn:2018hqm}. We include the contributions from uranium-235 and plutonium-239~\footnote{JEFF~3.3 does not contain fission yields for 14\,MeV neutrons for this isotope and thus we use ENDF/B-VIII.1 fission yields. A general comparison between neutrino fluxes obtained from JEFF and ENDF fission yields indicates agreement for the IBD rate at the 1--2\% level.}, and we compute their contributions as outlined above; our results agree well with a similar calculation published previously~\cite{Distel:2024qzg}.


We assume a segmented antineutrino detector of a mass of 1 metric ton comprised of solid organic scintillator similar to the miniCHANDLER detector~\cite{hag20,Rodrigues:2025hqk}. These solid plastic detectors can be operated remotely and with a minimum of maintenance. Note that the energy visible in this type of detector for an IBD event, the so-called prompt energy, is given by
\[
E_\mathrm{prompt} = E_{\bar\nu} - 1.806\,\mathrm{MeV} + 2\,\mathrm{m}_e = E_{\bar\nu} - 0.784\,\mathrm{MeV,}
\]

where the term $2\,\mathrm{m}_e = 1.022\,$MeV stems from the energy released when the positron annihilates with one of the detector electrons. We use a signal efficiency of 25\%~\cite{PROSPECT:2020sxr,PROSPECT:2021jey} and a distance of 25\,m. We assume that no significant number of fusion neutrons can reach the detector directly, i.e., their contribution to the background rate is negligible relative to all other backgrounds. For estimating the required level of shielding, we stipulate that we want no more than one fusion neutron per day and~$\mathrm{m}^2$, the approximate surface area of the detector.  To achieve this goal, the equivalent of about 6--7\,m of concrete shielding are required for the fusion neutron emission rate of $5.3\times10^{20}$ per second, the given detector distance, and the macroscopic removal cross section for 14\,MeV neutrons in concrete of 0.07--0.08\,cm$^{-1}$ \cite{Glasstone}. The specific target number for the tolerable neutron flux is not important since 1\,m of concrete provides a 3,000-fold attenuation. Much, maybe all, of this shielding will be provided by the breeding blanket and the biological shield of the reactor. We include cosmogenic backgrounds modeled after Refs.~\cite{PROSPECT:2020sxr,PROSPECT:2021jey}. The required fusion neutron shielding corresponds to about 14--17 meter water equivalent. If we assume a similar amount of shielding against cosmic rays  a significant reduction of cosmogenic backgrounds will result~\cite{Cogswell:2021lla}. We will use a total of 20 events per day per ton of detector in the prompt energy range of 1--9\,MeV following a $1/E_\mathrm{prompt}$-spectrum. These assumptions present a conservative detector performance and background model for a detector operating under these conditions. We also include the backgrounds from activation products in the blanket as computed earlier.

\begin{figure}[hbt] \centering 
\includegraphics[width=110mm]{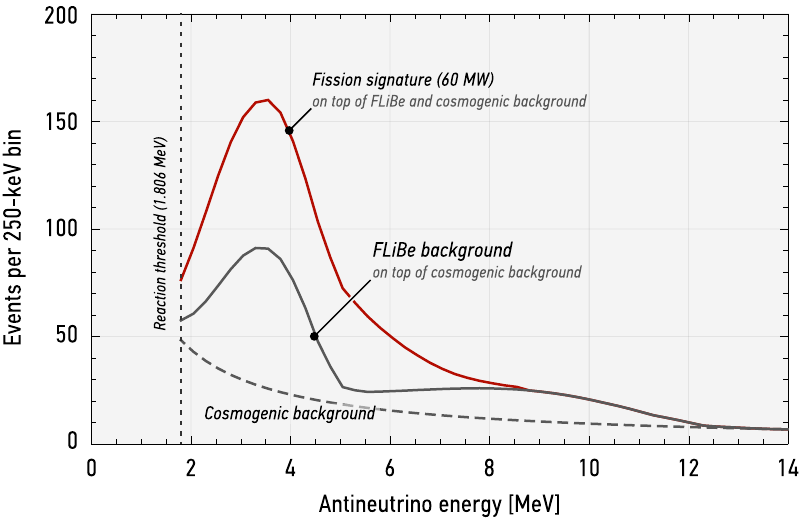}
\caption{Antineutrino event spectrum from fission events on top of the neutron-activation and cosmogenic background in a one-ton detector located 25\,m from the reactor using a FLiBe blanket. Data are collected over a 30-day period. All neutron activation products are included for 1500\,MW of fusion power.}
\label{fig:spectrum}
\end{figure}

The resulting event spectrum is shown in Fig.~\ref{fig:spectrum} assuming the lower fission source term of 60\,MW.
We can clearly see that the fission signal corresponding to the production of about 8\,kg of plutonium is well above both the cosmogenic and the reactor-produced backgrounds even for the FLiBe design. The shape of the fission signal is distinct from the shape of both backgrounds and, thus, can be unambiguously identified as a sign of plutonium production at a rate of approximately 1\,SQ per 30 days (or 330\,g/day).

Higher plutonium production rates would result in even easier to identify signals. On the other hand a lower plutonium production rate would lead to  a decrease of the fission signal. If we assume that the lower production rate is due to a lower concentration of fertile material in the blanket but the fusion power is kept the same, then also the reactor backgrounds would stay the same. The cosmogenic backgrounds will stay the same in any case. In combination, therefore a lower plutonium production rate effectively reduces the signal while the backgrounds remain unchanged. The question thus becomes, what is the lowest production rate that still can be reliably identified?

To answer this question, we use the event rates to construct a binned (in prompt energy) Poisson-likelihood function and use that to compute detection sensitivity. We use a likelihood-ratio hypothesis test where the null hypothesis is that no fissile material breeding takes place and we evaluate the smallest fission power, $P_\mathrm{fiss}$, for which we can reject the null hypothesis. We choose the critical value for the likelihood ratio such that we achieve a 95\% detection probability and a 1\% false positive rate. This critical value is determined under the assumption of normal distributed errors, which is well fulfilled. We assume a 10\%  normalization of the cosmogenic background when finding the maximum of the likelihood and we include also a 10\% error on the fusion power as well and profile over the corresponding nuisance parameter with associated pull terms to account for their respective errors.

\begin{table}[hbt]
\centering
\begin{tabular}{rrrr}
\hline
 Time  & Power limit  & Production rate  & Total mass \\
 days  &         MW &          g/day &       kg \\
\hline
   30 & 9.3\,/\,15.1 &  31.1\,/\,80.5 &   0.9\,/\,2.4 \\
   90 &  5.9\,/\,9.8 &  19.8\,/\,52.3 &   1.8\,/\,4.7 \\
  180 &  4.3\,/\,7.2 &  14.4\,/\,38.4 &   2.6\,/\,6.9 \\
  360 &  3.1\,/\,5.2 &  10.3\,/\,27.6 &   3.7\,/\,9.9 \\
 1800 &  1.4\,/\,2.4 &   4.7\,/\,12.5 &  8.5\,/\,22.5 \\
\hline
\end{tabular}
\medskip
\caption{Upper limits derived from neutrino observations for a detection probability of 99\% and false positive rate of 1\%. For number pairs, the first number is for the DCLL design, the second number is for the FLiBe design. Time is the observation period, and the power limit denotes the limit on the fission power. This power limit translates into a plutonium production rate limit in the third column, and multiplying this number with the number of days in the first column yields the limit on the total mass of plutonium produced shown in the last column.}
\label{tab:neutrino}
\end{table}

In Tab.~\ref{tab:neutrino} we show the result of this analysis for a range of observation periods ranging from 1 months to 5 years, where we assume that no fission neutrinos have been detected. The plutonium production rate is assumed to strictly proportional to the fission power and that the proportionality constant stays unchanged for a range of fertile material mass,  see {\it e.g.} Fig.~10 of Ref.~\cite{bal25}. The neutrino detector will provide an upper bound on the fission power and this bound scales with $1/\sqrt{t}$, where $t$ is the observation time. The differences of the values between the DCLL and FLiBe designs are due to two factors that push in the same direction: First, the FLiBe system has a significant neutrino background from fluorine-20 and lithium-8, that either is absent or much smaller, respectively, for the DCLL system; second, the higher conversion efficiency of the FLiBe system, {\it i.e.} per MW of fission power about 59\% more plutonium is produced. We also observe that the limit on the production rate improves with time, but the limit on the more relevant total mass decreases with time. This is not surprising, the former scales as $1/\sqrt{t}$ and the latter is simply the product of the former and time $t$ and thus scales as $1/\sqrt{t}\times t=\sqrt{t}$. Therefore there is a longest time that allows to set a limit on total plutonium mass of 8\,kg (1SQ) or less even if no fission neutrinos are observed. These times are 1,609 and 237\,days for the DCLL and FLiBe designs, respectively. These times represent the longest time periods where a neutrino measurement \emph{by itself} can ensure that no significant quantity of plutonium has been produced. 


\section{Conclusion}

Fusion energy systems show great promise as a future source of carbon-free electricity. These systems can in principle also be covertly used, however, for the production of weapon-usable fissile material. Demonstrating methods to assure the timely detection of such a misuse scenario would be extremely valuable as part of the efforts to guarantee the peaceful use of fusion energy systems.

The approach proposed here may offer an efficient and non-intrusive strategy to address this challenge. Our results confirm that even a modestly-sized antineutrino detector should be able to detect the presence of fertile material during operation in a reliable and timely manner. In particular, the production of one significant quantity (8~kg) of plutonium in one month provides a strong and easily detectable signature.

Naturally, lower plutonium production rates would be more challenging to detect as the fission signature gradually decreases compared to the antineutrino background. Scaling our reference results to examine such scenarios suggests that a production rate of one significant quantity stretched out over a period of one year would still be detectable with very high confidence for a reactor that uses a dual-coolant lead-lithium (DCLL) blanket. The detection task is more challenging for the reactor using the molten-salt (FLiBe) blanket, primarily, due to increased background emissions associated with fluorine-20 decay. 

Further work is required to optimize such a detector system and its shielding, and a wider range of blanket designs should be studied. Similarly, additional covert production scenarios could also consider thorium as a fertile material, in which case the fission signature would be weaker compared to our uranium reference case. The results reported here suggest that such an effort is well justified. Clearly, there is an opportunity for the fusion and the safeguards communities to explore a combination of monitoring and verification approaches that could complement each other and provide strong safeguards against misuse of future fusion energy systems.

~


\section*{Acknowledgments}

The authors thank Owen Webster of North Carolina State University for asking the initial question as to whether fusion power systems could be detected from their neutrino emission, which led us to investigate the question as to whether misuse of such systems could be detected. This work was supported by the U.S. Department of Energy National Nuclear Security Administration's Enabling Capabilities in Technology (TecH) Consortium under award number DE-NA0004197 and the Consortium for Monitoring, Technology, and Verification (MTV) under award number DE-NA0003920. It was also supported by the U.S. Department of Energy Office of Science under award number DE-SC0020262 and under contract number DE-AC02-09CH11466.

The United States Government retains a non-exclusive, paid-up, irrevocable, world-wide license to publish or reproduce the published form of this manuscript, or allow others to do so, for United States Government purposes.



\bibliographystyle{plain}
\bibliography{main}

\end{document}